\documentclass[twocolumn]{elsart}
\usepackage{amssymb}   
\usepackage{epsfig}

\psfigdriver{dvips}             

\begin{document}

\begin{frontmatter}

\title{Observation of Ising-like critical fluctuations in frustrated Josephson junction arrays with modulated coupling energies}
\author{J. Affolter, M. Tesei, Ch. Leemann, P. Martinoli}
\address{Institut de Physique, Universit\'{e} de Neuch\^{a}tel, 2000 Neuch\^{a}tel, Switzerland}
\author{H. Pastoriza}
\address{Centro Atomico Bariloche, 8400 San Carlos de Bariloche,
Rio Negro, Argentina}

\begin{abstract} We report the results of ac sheet conductance
measurements performed on fully frustrated square arrays of
Josephson junctions whose coupling energy is periodically
modulated in one of the principal lattice directions. Such systems
are predicted to exhibit two distinct transitions: a
low-temperature Ising-like transition triggered by the
proliferation of domain walls and a high-temperature transition
driven by the vortex unbinding mechanism of the
Beresinskii-Kosterlitz-Thouless (BKT) theory. Both the superfluid
and dissipative components of the conductance are found to exhibit
features which unambiguously demonstrate the existence of a double
transition whose properties are consistent with the Ising-BKT
scenario.


\end{abstract}

\end{frontmatter}

Two-dimensional arrays of Josephson junctions (JJA) exposed to a
perpendicular magnetic field provide the opportunity to study the
influence of a tunable level of frustration in a variety of
topologies ranging from periodic to random structures, including
quasiperiodic and fractal lattices. Such systems are a physical
realization of the frustrated classical XY model where the degree
of frustration is governed by a parameter f expressing the
magnetic flux threading an elementary cell of the array in units
of the superconducting flux quantum.

While the nature of the superconducting transition of a Josephson
junction array (JJA) at arbitrary frustrations is still not well
understood, the critical behavior of square JJAs at full
frustration (f=1/2) has been widely investigated theoretically and
with numerical simulations. Because of the "checkerboard"
structure of the ground state \cite{Teitel}, two symmetries are
relevant in determining the critical behavior of a fully
frustrated JJA: the continuous U(1) rotational symmetry and the
discrete $Z_{2}$ chiral symmetry \cite{Teitel,Halsey,Korshunov1}.
The phase transition resulting by breaking U(1) is driven, at a
temperature $T_{BKT}$, by the vortex unbinding mechanism predicted
by the Beresinskii-Kosterlitz-Thouless (BKT) theory, whereas the
transition associated with $Z_{2}$ is triggered by the
proliferation of Ising-like domain walls at a temperature $T_{I}$.
The question of whether these two transitions are disinct
\cite{Olsson} or merge into a single transition belonging to a new
universality class \cite{Lee} has been controversial for a long
time. Quite recently Korshunov \cite{Korshunov2} has shown that
the first scenario, with $T_{I}>T_{BKT}$, is the only possible
one. Although distinct, the two transitions are, however,
very close to each other, making 
experiments conceived to explore their exact nature rather
difficult. From an experimental point of view, the situation is
more favorable in frustrated arrays whose coupling energy $E_{J}$
is periodically modulated in one of the principal lattice
directions. Their critical behavior was first studied with Monte
Carlo simulations by Berge et al. \cite{Berge} and, subsequently,
by Eikmans et al. \cite{Eikmans} using a Coulomb gas approach. In
arrays with a sufficiently strong $E_{J}$-modulation the two
transitions are predicted to be well separated (with
$T_{I}<T_{BKT}$) and thus accessible to experimental observation.
In this brief report we present preliminary results of ac sheet
conductance measurements performed on JJAs with modulated
couplings at f=1/2 which unambiguously demonstrate the existence
of a double transition with features consistent with the Ising-KT
scenario.

\begin{figure}[h]
 \begin{center}
 \includegraphics{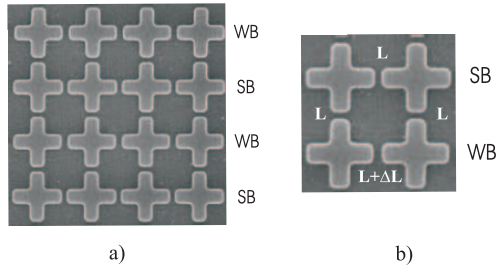}
 \caption{\label{sample}a) SEM picture of a portion of the array
 showing Pb crosses on a Cu ground plane. The lattice parameter is $a=8\mu m$.
 The "weak" JJs  compose the first and third lines (WB), whereas the "strong"
 junctions (SB) compose the second and fourth lines.
 b) Picture of one square plaquette of the array showing the three
 "strong" junctions (L=$0.8\mu m$) and the lengthened junction (L+$\Delta$L=$0.9\mu m$).}
 \end{center}
\end{figure}

The experiments were carried out on square arrays of
proximity-effect coupled SNS
(superconducting-normal-superconducting) Josephson junctions
consisting of $\sim 10^{6}$ Pb superconducting (S) islands forming
a square lattice on a normal (N) Cu layer. To periodically
modulate $E_{J}$, the N-bridges of the junctions located on
alternating rows were slightly lengthened in order to decrease
their coupling energy with respect to that of all the other
junctions in the array (see Fig. \ref{sample}). The lengths of the
N-metal gaps of the array shown in Fig. \ref{sample} are $0.8\mu
m$ for the strong bonds (SB) and $0.92\mu m$ for the weak bonds
(WB). The transition temperature of the S-islands was $T_{CS}=7.02
K$ and the normal metal coherence length $\xi_{N}(T_{CS})\cong 80
nm$. Kinetic inductance measurements performed on the unfrustrated
(f=0) array  \cite{Leemann} allow to estimate the ratio $\eta$
between the coupling energies of the weak and the strong bonds. We
find $\eta\cong 0.4$ in the temperature range of interest. The
sheet resistance of the array in the normal state was $R_{N}\cong
3 m\Omega$.

To explore the properties of the array in the critical region, we
measured its complex ac sheet conductance $G(\omega, T)$ at f=1/2
using a SQUID-operated two-coil mutual inductance technique
\cite{Jeanneret}, which allows to probe critical fluctuations over
a frequency range covering more than 5 decades (0.1 Hz - 10 kHz).
At the temperatures of interest ($T\cong 5 K$) we estimate that
the vortex diffusion length ($r_{\omega}/a\sim (14 R_{N} k_{B} T/
\omega \phi_{0}^{2})^{1/2}$) is about $10^{3}$ lattice constants
at the lowest accessible frequencies. Thus, at these very large
length scales, our low-frequency conductance measurements should
reflect the response of the array in the quasi-static
thermodynamic limit. In this regime therefore the inverse sheet
inductance $L_{\square}^{-1}=\omega Im(G)$, which is proportional
to the areal superfluid density and measures the degree of
superconducting phase coherence in the system, is directly
comparable with Monte Carlo simulations of the array helicity
modulus \cite{Eikmans}.

\begin{figure}[h]
 \begin{center}
 \includegraphics{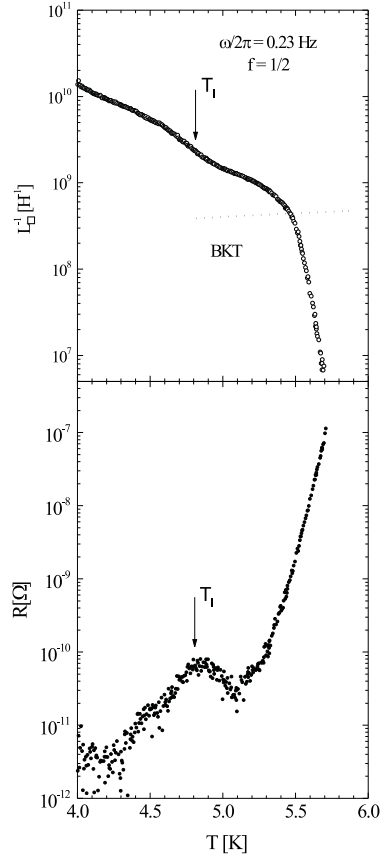}
 \caption{\label{LsqR(T)}Inverse sheet inductance ($L_{\square}^{-1}$)
 and resistance (R) vs temperature at full frustration f=1/2.
 The softening of the superconducting phase coherence at $T_{I}$
 is evidenced by a depression of $L_{\square}^{-1}$ and a corresponding
 shallow peak in R. The total disappearance of phase coherence and the
 rapid increase of dissipation are the signatures of the BKT
 transition.
 Dotted line: BKT prediction for $L_{\square}^{-1}(T_{BKT})$.}
 \end{center}
\end{figure}

In Fig. \ref{LsqR(T)} both the superfluid ($L_{\square}^{-1}$) and
the dissipative (R=Re(1/G)) components of the conductance
extracted from the linear response of the fully frustrated (f=1/2)
array at an excitation frequency of 0.23 Hz are shown as a
function of temperature on log-lin plots. At $T\cong 4.8 K$ there
is a slight depression in $L_{\square}^{-1}(T)$ accompanied by a
peak in dissipation. Relying on the Coulomb gas analogy
\cite{Eikmans}, we identify these features as the signatures of
the "antiferroelectric-paraelectric" Ising transition triggered by
the proliferation of domain walls in the system of dipoles created
by the attractive interaction, across the weak bonds, of
half-integer charges. This interpretation is supported by the
observation that, in terms of the reduced temperature $\tau =
k_{B}T/E_{J}(T)$, the transition takes place at $\tau_{I}=0.08$,
in good agreement with the value ($\tau_{I}= 0.07$) predicted by
the phase diagram of Ref.7 for $\eta\cong 0.4$. At higher
temperatures the superfluid component drops dramatically,
signaling the suppression of global superconducting phase
coherence, at a temperature $T_{BKT}\cong 5.5 K$, which is found
to be consistent, as shown by the dotted line in Fig.
\ref{LsqR(T)}, with the universal BKT prediction
$L_{\square}^{-1}(T_{BKT}) = \left( 8\pi/\phi_{0}^{2}\right)
k_{B}T_{BKT}$.

According to the Coulomb gas analysis \cite{Eikmans}, the kink
structure in $L_{\square}^{-1}(T)$ at $T_{I}$ reflects a
logarithmic anomaly in the susceptibility of the Ising-like system
of oriented dipoles. More precisely, the temperature derivative
$dL_{\square}^{-1}/dt$ (where $t=1-T/T_{I}$) should exhibit, at
$T_{I}$, a $ln|t|$ divergence, which can be studied in great
detail by varying the length scale (i.e. the driving frequency
$\omega$) at which one is probing the JJA. The results of these
investigations and, more generally, of the very interesting
frequency dependence of the dynamic response in the critical
Ising-BKT region will be published elsewhere.

We would like to thank S.E. Korshunov for useful discussions. This
work was supported by the Swiss National Science Foundation, the
Swiss Federal Office for Education and Science within the
framework of the TMR network "Superconducting Nanocircuits" of the
European Union and  the Vortex Program of the European Science
Foundation.


\end{document}